\documentclass[a4paper]{spie}  

\usepackage{amsmath,amsfonts,amssymb}
\usepackage{graphicx}
\usepackage[colorlinks=true, allcolors=blue]{hyperref}
\usepackage{enumitem}
\setlist{nosep} 

\title{The MICADO first light imager for the ELT: MISTHIC simulation pipeline for the high contrast mode of MICADO}

\author[a]{E. Huby}
\author[a]{P. Baudoz}
\author[a]{F. Vidal}
\author[a]{H. Baran}
\author[a]{Y. Cl\'enet}
\author[b]{R. Davies}
\affil[a]{LESIA, Observatoire de Paris, Universit\'e PSL, CNRS, Sorbonne Universit\'e, Universit\'e Paris Cit\'e, 5 place Jules Janssen, 92195 Meudon, France}
\affil[b]{Max-Planck-Institut für extraterrestrische Physik, Garching, Germany}

\authorinfo{Further author information: (Send correspondence to E. H. and P. B.)\\E. H.: elsa.huby@obspm.fr ; P. B.: pierre.baudoz@obspm.fr}

\begin{document}
\maketitle

\begin{abstract}
We present the Python pipeline that was developed to simulate the high contrast mode images of the MICADO instrument. This mode will comprise three classical Lyot coronagraphs with different occulting spot sizes, one vector apodized phase plate, and two sparse aperture masks. One critical aspect of these modes lies in their sensitivity to aberrations, requiring careful integration of non common path aberrations, as well as turbulent adaptive optics (AO) residuals. Besides, they will be operated following a specific observing strategy based on pupil tracking mode. For these reasons, we have developed the MICADO SimulaTor for HIgh Contrast (MISTHIC) pipeline to simulate realistic image cubes and derive the expected performance of these modes. Several sources of aberrations can be included: turbulent AO residuals, but also static or rotating aberrations, amplitude aberrations, simple Zernike screens… Including MICADO's specific features, such as the absence of atmospheric dispersion prior to the coronagraphic focal plane mask, our intent is to make this pipeline available to the community. This tool can be used to prepare scientific observations with the high contrast mode of MICADO, by predicting the performance to be expected within the most realistic assumptions.
\end{abstract}


\section{INTRODUCTION}
\label{sec:intro}  

\begin{figure}
    \centering
    \includegraphics[width=0.8\linewidth]{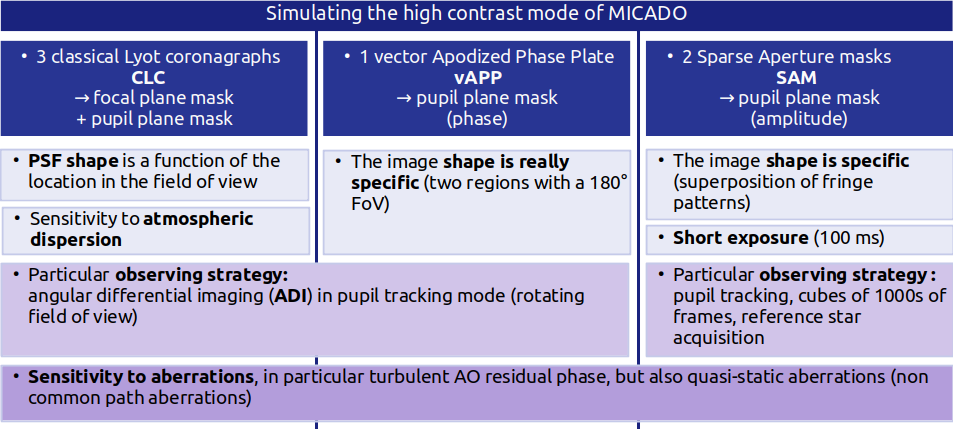}
    \caption{Specificities of each observing mode of the high contrast mode of MICADO, to be considered for realistic simulations.}
    \label{fig:modes}
\end{figure}

We have developed the \texttt{MIcado SimulaTor for HIgh Contrast} (\texttt{MISTHIC}), detailed in Section\,\ref{sec:workflow}, to predict the expected performance of the high contrast mode of MICADO\cite{Davies2016} under realistic conditions. The pipeline has been designed to simulate images obtained with the three different observing modes offered in the high contrast mode of the MICADO instrument:
\begin{itemize}
    \item classical Lyot coronagraphs, consisting in the combination of an occulting mask in the focal plane and a Lyot stop in the pupil plane\cite{Perrot2018}. There will be three available occulting masks, with the following radii: 15\,mas (CLC15), 25\,mas (CLC25) and 50\,mas (CLC50). The Lyot stop has a diameter of 86.2\% of the nominal ELT pupil diameter (38.542\,m), a central obstruction of 39.2\% and spider arms oversized by a factor of 3.  Examples of CLC simulated images are shown in Section\,\ref{sec:clc}.
    \item Vector Apodized Phase Plate (VAPP), a phase mask located in a pupil plane, which reshapes the incoming beam diffracted in the subsequent focal plane to carve a dark region around the central star, where the starlight is rejected elsewhere in the field of view\cite{Codona2006}. The grating technique allows splitting the beam into two, in order to access a deep contrast performance in a defined area on each side of the star image\cite{Snik2012,Doelman2021}. 
    \item Sparse Aperture Masks: in the presence of substantial turbulent phase residuals, interferometry can be powerful in calibrating out phase variations across the pupil. There are two available masks with 9 and 18 holes (see Huby et al., these proceedings). Example of SAM simulated images are shown in Section\,\ref{sec:sam}.
\end{itemize}
Each of these modes presents peculiarities that have to be taken into account, regarding the image shape or acquisition procedure, as summarized in Figure\,\ref{fig:modes}. CLC images vary depending on the location in the field of view, particularly at short separations from the coronagraphic mask center. In addition, depending on the spectral bandwidth, atmospheric dispersion can impact the final image, since the atmospheric dispersion corrector is located downstream of the coronagraphic focal plane. One critical aspect is that the performance of each observing mode heavily relies on the wavefront quality and stability. Optical aberrations such as non common path aberrations and turbulent phase residuals must be carefully integrated. Moreover, such observations will be conducted with a specific strategy requiring (i) pupil stabilization to enable a PSF as stable as possible (pupil tracking mode, inducing a rotating field of view) and (ii) the acquisition of many short (a few seconds) or very short (100\,ms) exposure frames to enable the calibration of the aberrations in post-processing. As a consequence, dedicated simulations are required to include any type of aberrations and predict their impact on the images and on the expected performance under realistic conditions.

Additionally, \texttt{MISTHIC} can be used to simulate images obtained with the spectroscopic mode of MICADO, a mode currently under investigation to characterize exoplanets\cite{PalmaBifani2023}. This case is described in Section\,\ref{sec:spectro}

\section{SIMULATION PIPELINE WORKFLOW}
\label{sec:workflow}

The whole \texttt{MISTHIC} code is based on an initial \texttt{IDL} version written by Pierre Baudoz, and is now contained in a Python package. It includes a main function to run the simulations, and secondary functions organized in modules. Light propagation from pupil to focal planes are modeled thanks to fast Fourier transforms (FFT). In the case of the occulting spots for classical Lyot coronagraph, the simulation is implemented following the semi-analytical method\cite{Soummer2007}: only the electric field falling on the area of the occulting spot is necessary to estimate the electric field at the Lyot plane. It can indeed be written as the input pupil field subtracted from the field corresponding to the field propagating from the focal plane occulting mask, which is actually blocked. This means that the electric field in the focal plane needs only to be computed in a very small field of view (from 30 to 100mas, depending on the mask). To that purpose, a matrix discrete Fourier Transform\cite{Soummer2007} (MFT, as implemented in the \texttt{poppy} package\footnote{https://github.com/spacetelescope/poppy}) is used to compute the electric field in this small area at a high sampling rate. In other cases where a decent sampling of the focal plane is required (e.g. vortex or four quadrant phase mask), MFT is also applied. Finally, the electric field propagating from the last pupil plane (Lyot plane) is computed thanks to FFTW\footnote{https://www.fftw.org/} implemented in Python \footnote{https://github.com/pyFFTW/pyFFTW}. Zero padding, necessary to adjust the sampling in the final detector image, is implemented in two steps: first computing the FFT in the x direction, then cropping the output along the x axis to the desired field of view, prior to applying the padding and FFT in the y direction. 

\begin{figure}
    \centering
    \includegraphics[width=\linewidth]{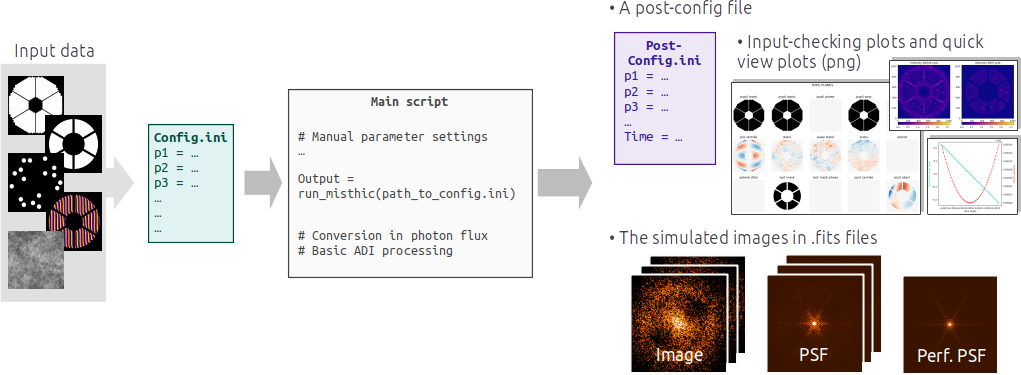}
    \caption{Global workflow of the \texttt{MISTHIC} simulation pipeline. Input parameters, including input data (pupil masks or aberration screens) are specified in a configuration file. The path to this \texttt{.ini} file is passed as argument of the main simulation function, \texttt{run\_misthic}. Outputs include the simulated image cubes saved as fits files, a copy of the configuration file for the record and some input-checking plots, which are visual representations used to verify the assumptions and parameters of our simulations.}
    \label{fig:workflow}
\end{figure}

The workflow for using the simulation tool is illustrated in Fig.\,\ref{fig:workflow} and consists in setting up a configuration file, and running a main function, \texttt{run\_misthic}, which takes the path of the configuration file as argument. This \texttt{.ini} file consists of a list of all parameters relevant to the simulation and is created using the \texttt{ConfigObj} module\footnote{https://pypi.org/project/configobj/}. All available input parameters and data are described in section \ref{sec:inputs}, and all outputs are detailed in \ref{sec:outputs}.

\subsection{Input parameters}
\label{sec:inputs}

The configuration file is organized in several sections. The \texttt{misthic\_configspec.ini} file specifies the expected type of each variable, and also sets default values in case it is missing in the user config file.

The \texttt{[simuconfig]} section includes general parameters such as the zenith distance (relevant to take the atmospheric refraction into account, or compute the parallactic angle), and observatory related parameters (telescope diameter, latitude). In the case of the ELT, the telescope diameter is defined as the nominal diameter according to ESO definition (i.e. $D_{tel}=38.542m$).

The \texttt{[waveconfig]} section defines the spectral width of the simulation and number of spectral channels that will emulate the broadband images. The single monochromatic images can be saved if specified. 

The \texttt{[aberrconfig]} section is a list of all aberrations that can be included, all specified as optical path difference, in microns:
\begin{itemize}
    \item aberrations defined either as a set of Zernike polynomial coefficients, or alternatively loaded from a fits file;
    \item amplitude aberrations loaded from a fits file;
    \item static aberrations randomly generated from a power spectral density decreasing with the spatial frequency as $f^{-2}$ (or loaded from a fits file);
    \item rotating aberrations following the parallactic angle variation, randomly generated from a PSD (or loaded from a fits file);
    \item turbulent phase residuals after AO correction, with the possibility to define the temporal sampling (see Section \ref{sec:clc}) and to specify the removal of the global slope of the phase (tip-tilt removal);    
    \item post-coronagraph aberrations defined either as a set of Zernike polynomial coefficients or loaded from a fits file.
\end{itemize}

The parameters relevant to computing the spectral deviation due to atmospheric refraction are also specified in this section. As it will be implemented in the MICADO instrument, it is possible to simulate the atmospheric refraction affecting the coronagraphic focal plane and apply the expected compensation with a (perfect) atmospheric dispersion corrector.

Finally the \texttt{[coroconfig]} section sets the type of imaging mode to be simulated: occulter mode (for CLC), VAPP or no focal plane mask (for SAM), but other coronagraphic masks like the vortex mask or the four quadrant phase mask can be implemented as well. The entrance pupil plane can be loaded from a fits file, or generated from geometrical parameters (outer diameter and central obstruction sizes). The same applies for the Lyot stop specification. In addition, Lyot stop misalignment can be simulated (shift or drift).

\subsection{Output products}
\label{sec:outputs}

\paragraph{Main outputs.} As illustrated in Fig.\,\ref{fig:workflow}, three fits files are saved as output of the simulation: 
\begin{itemize}
    \item the file ending with ``*\_image\_cube.fits'' contains the image cube simulated with the coronagraphic mask and all specified aberrations;
    \item the file ending with ``*\_psf\_cube.fits'' contains the PSF obtain under the same configuration as the coronagraphic images (aberrations, amplitude apodization and Lyot stop), except for the focal plane coronagraphic mask;
    \item the file ending with ``*\_perfect\_psf.fits'' contains a single-frame image of a perfect PSF, computed using the input pupil without a coronographic mask, Lyot stop, or aberrations. This PSF is used for normalization of all output images by its maximal value.
\end{itemize} 

Also outputted is a table with the parallactic angle values corresponding to each image simulated in the cube (based on the zenith distance and specified exposure time). These values are useful for ADI processing.

At the completion of each simulation, a copy of the configuration file containing the simulation parameters is saved in the same output directory as the data cubes. All simulation parameters are thus recorded, and can later be used as input to run the simulation again if needed.

\paragraph{Optional outputs} Figures can be plotted and saved in png format. In particular, the pupil plane plot can help checking the inputs of the simulation. Optionally, images before and after the Lyot stop can be saved as fits files if desired. Also, a polychromatic cube with the images simulated in every spectral channel can be saved in fits file too (only if the number of images to simulate is set to a single frame).

\section{Example results}

\subsection{CLC simulation with exoplanets}
\label{sec:clc}

In order to assess the performance of the CLC observing mode, extensive simulations have been run to emulate the detection of an exoplanet in the vicinity of the star. We expect the residual turbulent phase to define the sensitivity limit or the observations. Turbulent phase screens as filtered by the AO system are thus an essential input of the simulations. Such simulations were performed using the GPU-based \texttt{COMPASS} simulator \cite{Gratadour2014,Vidal2022}, emulating the MICADO AO system running at 500\,Hz, for different seeing conditions (best seeing Q1 of 0.49", worst seeing Q4 of 1.19" and median seeing of 0.70"). At the output, one out of 20 frames are saved, generating residual phase screens worth of 30\,min of observations, at the sampling of 25\,Hz, i.e. a total of 45\,000\,optical path difference screens.

\begin{figure}
    \centering
    \includegraphics[width=0.8\linewidth]{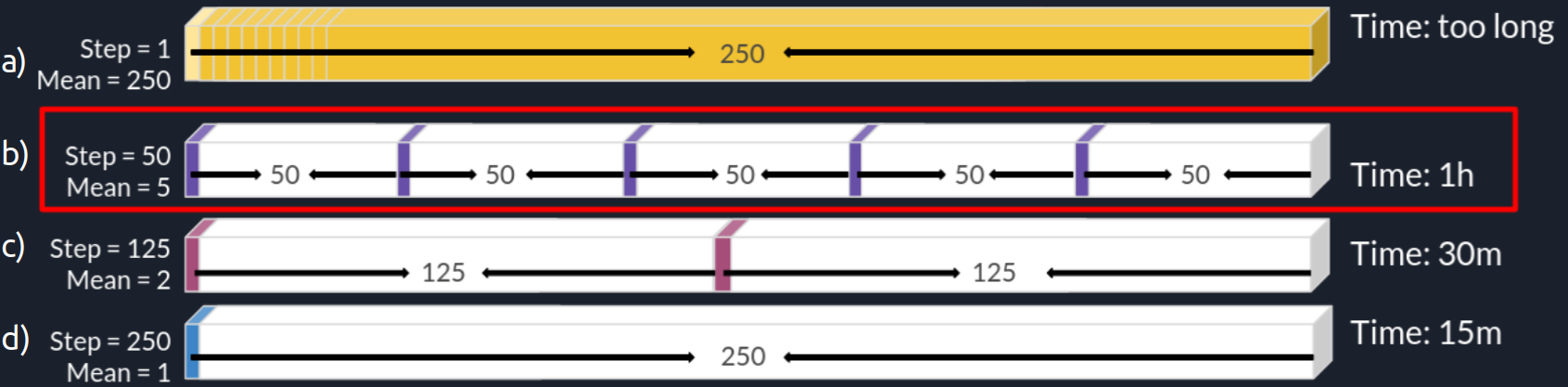}
    \caption{Schematic representation of the temporal sampling explored to account for turbulence residuals in the simulation of the planet PSF. Every line represents the 250\,frames corresponding to 10\,sec (25\,Hz), while the frames to be averaged are highlighted with color: a) all 250\,frames (25\,Hz), b) one frame out of 50 (0.5\,Hz), c) one frame out of 125 (0.2\,Hz), d) one frame out of 250 (0.1\,Hz). On the right hand side are displayed the approximate computation time to simulate 180 such frames (typically on a laptop).}
    \label{fig:temp_sampling}
\end{figure}

To simulate images at a typical exposure time of 10\,sec, 250\,images need to be averaged to simulate every single frame (180 such frames in total). Since the image of the star, centered on the coronagraphic mask, is highly sensitive to aberrations and thus turbulence residuals, this high sampling was deemed necessary to simulate the coronagraphic image of the central star. However, the image of the exoplanet is not affected or mostly unaffected by the coronagraphic mask, and is much less sensitive to the temporal variations of the aberrations. The temporal sampling of the turbulence residuals can thus be reduced. We led a study to determine whether the temporal sampling could be reduced to 0.5, 0.2 or 0.1\,Hz, i.e. the number of frames to be averaged could be decreased to 5, 2, or 1 frame to simulate a 10\,sec image, as illustrated in Figure\,\ref{fig:temp_sampling}. This would save time, as we intended to simulate the planet image at multiple location in the field of view, for the three different CLC and at least for the three different spectral filters.

\begin{figure}
    \centering
    \includegraphics[width=\linewidth]{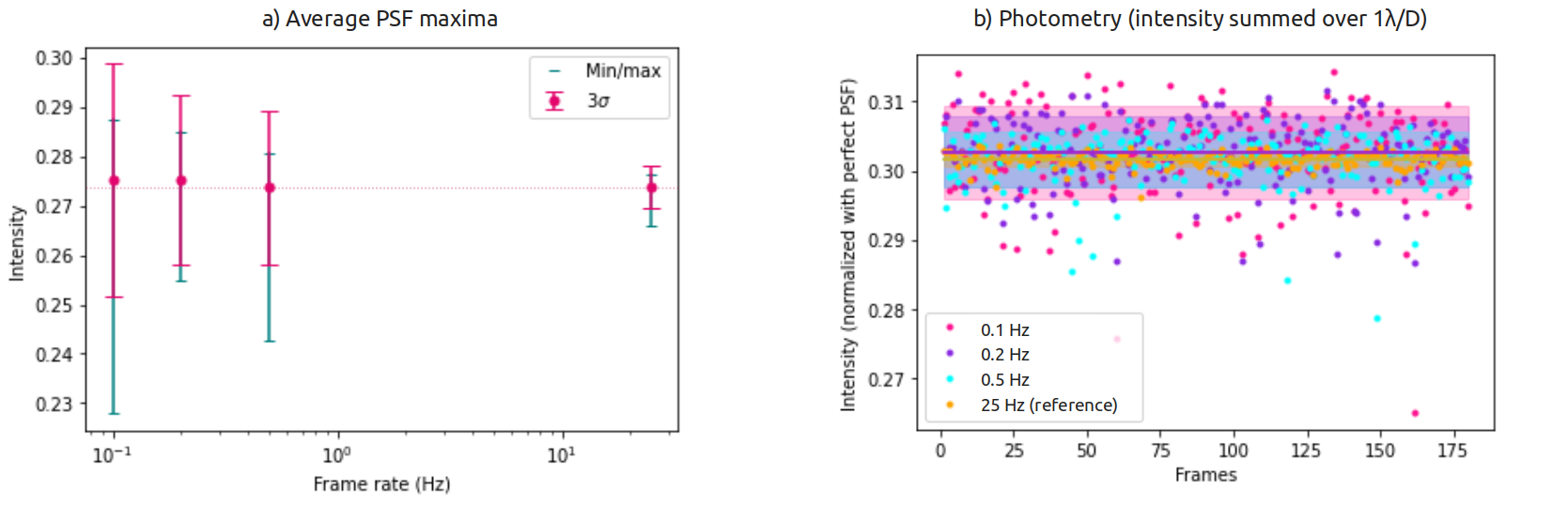}
    \caption{Comparison of the Strehl ratio and planet photometry in the 180 10\,sec-frame cube, computed with different temporal sampling: 25\,Hz, 0.5\,Hz, 0.2\,Hz, 0.1\,Hz. The simulations were run for CLC25 at 1.190$\mu m$ and best seeing conditions (Q1). The normalization is done on the perfect PSF without Lyot stop. Thus, the intensity values shown in these plots correspond to the Strehl ratio degraded by the transmission and PSF dilution induced by the Lyot diaphragm.}
    \label{fig:sampling_comp}
\end{figure}

\begin{figure}
    \centering
    \includegraphics[width=\linewidth]{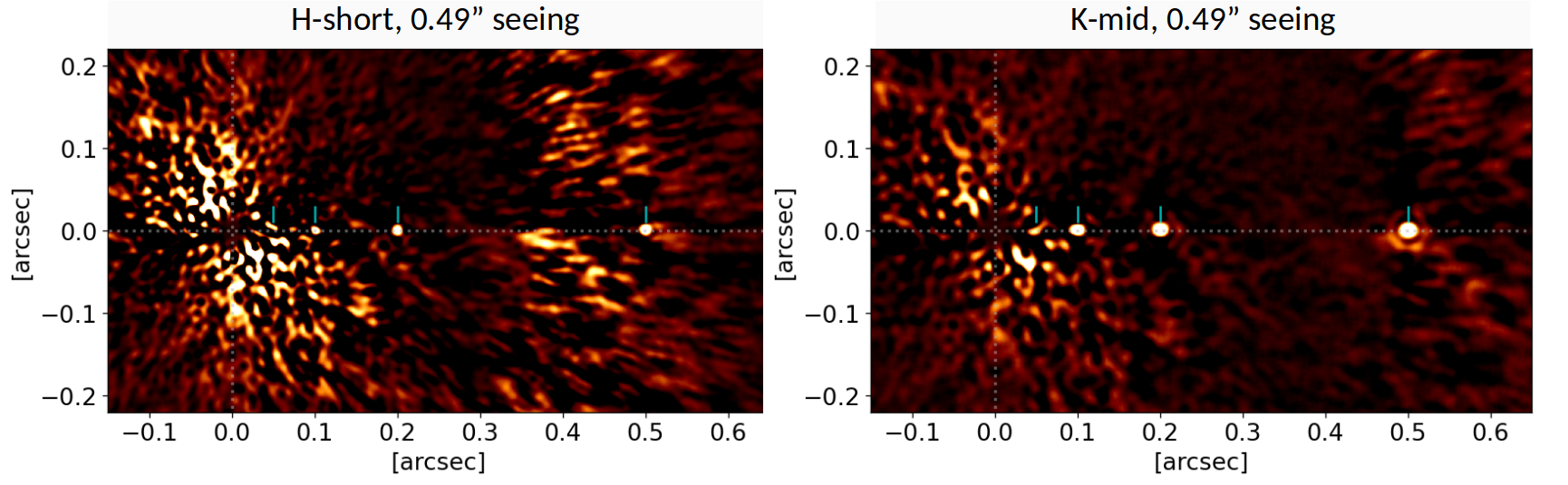}
    \caption{Final ADI-processed image for 30\,min observation sequences in H-short and K-mid filters with CLC25. The four simulated planets are highlighted with light blue lines at their different angular separations, and have the same intensity. The star behind the coronagraphic mask is located in (0,0) coordinates.}
    \label{fig:CLC25_results}
\end{figure}

The comparison between the different simulation regimes is based on the evaluation of the Strehl ratio along the 180 frames. On the left hand side of Figure \ref{fig:sampling_comp} are shown the average maximal pixel, along with the associated standard deviation, minimum and maximum values. On the right hand side of the figure is plotted the integrated flux in an area of 1$\lambda/D$ in diameter, reflecting a quantity closer to what would be done for a photometric study of the planetary companion. Both plots tend to show the same conclusion: decreasing the temporal sampling of the turbulent screens lead to a larger scatter of the Strehl ratio and integrated photometry, and there are a few more outliers towards lower values. While the presented graphs relate to CLC25 at 1.190$\mu$m (J band central wavelength) under the best seeing conditions (Q1), the study was also extended to the central wavelengths of H and K bands, and for median and worst seeing conditions (Q4), showing the same trend. Given these results, we considered that a sampling of 0.5\,Hz would be sufficient to obtain results that are, in average, close enough to the highest temporal sampling, for a significant gain in computation time.

As example results, Figure\,\ref{fig:CLC25_results} shows the coronagraphic images including planets at 800\,K orbiting a star similar to HR8799 ($M_H$=5.28, $M_K$=5.24, distance of 39.4\,pc), as detected in a 30\,min observation with CLC25 in H-short and K-mid filters, under the best seeing conditions. The planet spectrum model\cite{Charnay2018} corresponds to a 1\,$R_{Jup}$ radius, $log(g)$=4, met=0.32, CO\,ratio=0.1. A classical angular differential imaging processing was applied, with a parallactic angle variation of 20\,degrees around the transit of the star (minimal zenithal angle of 20\,deg). The planets stand out from the speckle noise for separations larger than 100\,mas, or 5\,AU at that distance.

\subsection{SAM simulation}
\label{sec:sam}
\begin{figure}
    \centering
    \includegraphics[width=\linewidth]{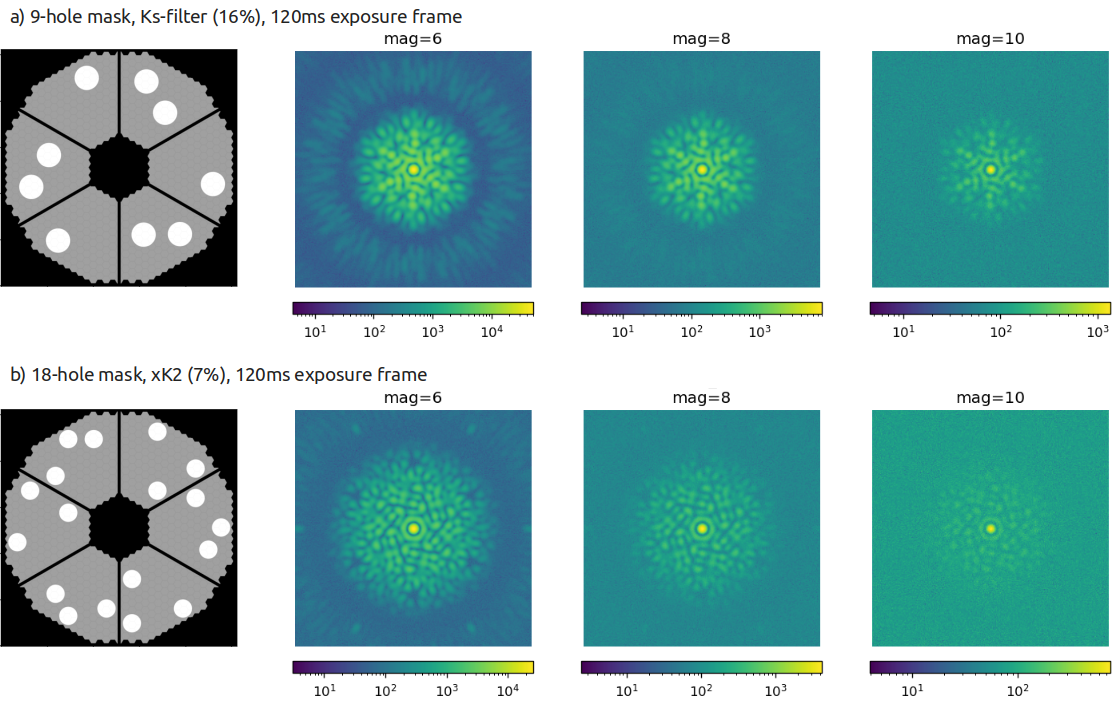}
    \caption{SAM image simulations with the a) 9-hole mask in broadband K filter and b) the 18-hole mask in medium band K filter, for different magnitudes and for an integration time of 120\,ms. The scale is logarithmic, in photon, and read-out-noise is included.}
    \label{fig:sam}
\end{figure}

Non-redundant imaging offers a way to enable the detection of features at very high angular separation \cite{Haniff1987}. While the performance in terms of contrast is currently limited to approximately $10^{-3}$, this sensitivity limit extends down to the diffraction limit of the telescope. Its implementation relies on the use of sparse aperture masks (SAM) located in a pupil plane. Such mask consists of an opaque mask drilled with holes positioned on a non-redundant configuration. This means that every pair of holes, called sub-pupils, forms a unique baseline vector. The analysis of the fringe pattern allows the estimation of the spatial coherence factors. The observed object is recovered either by a fit of the closure phases with a given model (e.g. star + companion), or by a fit of the complex visibilities to reconstruct an image. SAM images simulated with \texttt{MISTHIC} were used to design the two masks that will equip MICADO and to derive the expected performance (see Huby et al., these proceedings). Example images are shown in Figure\,\ref{fig:sam}.

\subsection{Spectroscopic mode}
\label{sec:spectro}

A last example of simulations performed with \texttt{MISTHIC} relates to the simulation of the long-slit spectroscopic mode of MICADO, investigated to characterize exoplanets \cite{PalmaBifani2023}. For the moment, simulations were run up to a spectral resolution of 1000 (500 spectral channels between 1.5 and 2.5$\mu$m). Images were simulated (without any coronagraph or mask), with careful consideration given to the wavelength sampling.

In terms of simulation, the change of wavelength implies a change of sampling in the final image. In the \texttt{MISTHIC} function, this sampling is tuned by adjusting the zero padding factor of the pupil image, to adjust the number of pixels within 1\,$\lambda/D$. For a desired sampling $S_\mathrm{det}$ at the minimal wavelength of the spectral band taken as the reference wavelength $\lambda_\mathrm{ref}$, the zero padding factor at wavelength $\lambda_\mathrm{ref}$ should be equal to $S_\mathrm{det}$. In other terms, the grid width should equal $S_\mathrm{det} \cdot D_\mathrm{pup} \cdot \lambda / \lambda_\mathrm{ref}$ at any wavelength $\lambda$, with $D_\mathrm{pup}$, the pupil diameter in pixels. The finest change in wavelength between $\lambda_1$ and $\lambda_2$ is simulated by adding 2 pixels to the previous grid width, which translates into a pixel difference of:

\begin{equation}
    S_\mathrm{det} \cdot D_\mathrm{pup} \cdot \frac{\lambda_2}{\lambda_\mathrm{ref}} = S_\mathrm{det} \cdot D_\mathrm{pup} \frac{\lambda_1}{\lambda_\mathrm{ref}} + 2.
\end{equation}

As a result, the smallest wavelength sampling is given by:
\begin{equation}
    \delta \lambda_\mathrm{min} = 2 \cdot \frac{\lambda_\mathrm{ref}}{S_\mathrm{det} \cdot D_\mathrm{pup}},
\end{equation}

\begin{figure}
    \centering
    \includegraphics[width=.8\linewidth]{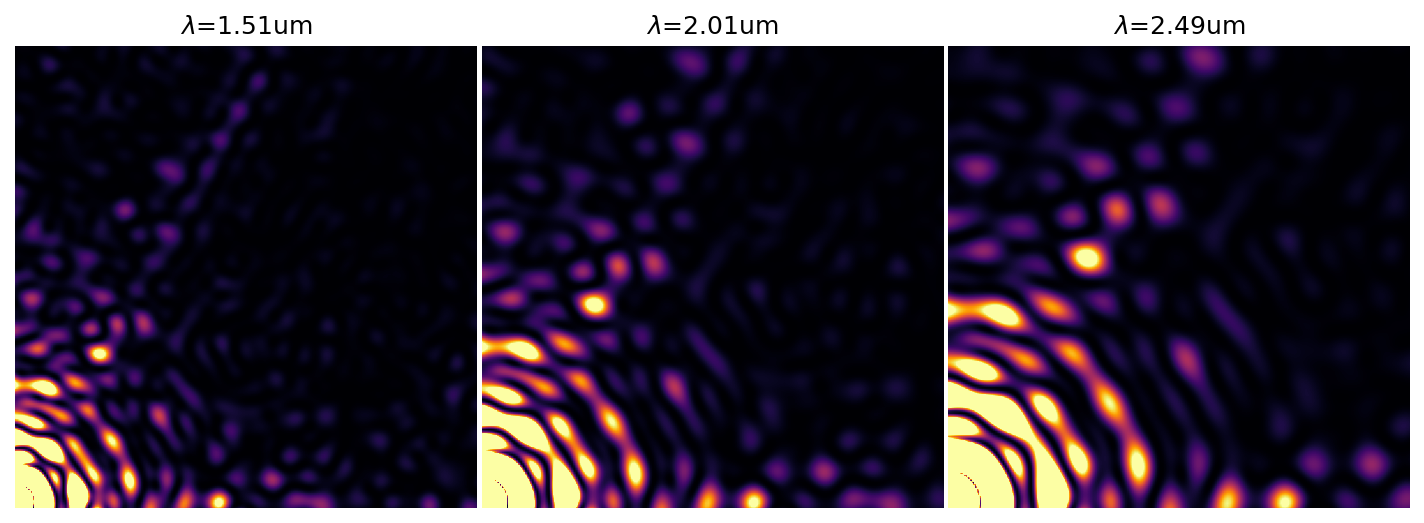}
    \caption{Spectroscopic image samples showing the change of scale between the wavelengths at the edge of the spectral bandwidth.}
    \label{fig:spectro}
\end{figure}

Hence, for $\lambda_\mathrm{ref}=1.51\mu m$, $D_\mathrm{pup}=1014.26$ (as defined as the nominal diameter of 38.542\,m in our ELT pupil image) and $S_\mathrm{det}=2$ (corresponding to a sampling of 4\,mas at $\lambda_\mathrm{ref}$), this leads to $\delta_\mathrm{min}=1.5$\,nm, or a maximal spectral resolution of about 1300. Increasing the detector sampling to 5.35 (i.e. a pixel scale of 1.5\,mas) leads to $\delta_\mathrm{min}=0.56$\,nm, or a maximal spectral resolution of about 3600. The first run of simulations was performed with $S_\mathrm{det} \sim 8$ enabling a spectral resolution of about 5400 at maximum. Sample images are shown in Figure\,\ref{fig:spectro}

\section{CONCLUSION}

The \texttt{MISTHIC} Python package has been developped to simulate images obtained with the high contrast mode of MICADO, including classical Lyot coronagraph, vector-APP and non-redundant masking images. While the code has not been particularly optimized for speed, the focus has been put on the flexibility to enable simulations using the same pipeline for the three different observing modes under realistic observing conditions. Future improvement could include the possibility to parallelize the computation to speed up the computation time. The pipeline will be available on a \texttt{github} repository in a near future.

\acknowledgments 

The MICADO project has benefited from the support of 1) the French Programme d’Investissement d’Avenir through the project F-CELT ANR-21-ESRE-0008, 2) the CNRS 80 PRIME program, 3) the CNRS INSU IR budget, 4) the Action Spécifique Haute Résolution Angulaire (ASHRA) of CNRS/INSU co-funded by CNES, 5) the Observatoire de Paris and 6) the Région Ile-de-France (DIM ACAV/ACAV+ and ORIGINES).

\bibliography{2024_SPIE_proc_misthic} 
\bibliographystyle{spiebib} 

\end{document}